\documentstyle{l-aa}
\begin{document}

  \thesaurus{  09.08.1;    
               11          
              (11.01.1;    
               11.09.4;    
               11.09.5;    
               11.19.2) }  

   \title{On the oxygen abundance determination in HII regions}

   \subtitle{High -- metallicity regions}

   \author{L.S. Pilyugin }

  \offprints{L.S.Pilyugin}

   \institute{ 
                 Main Astronomical Observatory
                 of National Academy of Sciences of Ukraine,
                 Goloseevo, 03680 Kiev-127, Ukraine, \\
                 (pilyugin@mao.kiev.ua) 
                 }
                 
   \date{Received ; accepted }

\maketitle

\markboth {L.S.Pilyugin: On the oxygen abundance determination in 
high-metallicity HII regions}{}

\begin{abstract}

This is our second paper devoted to the problem of line 
intensity -- oxygen abundance calibration starting from the idea of McGaugh 
(1991) that the strong oxygen lines ($[OII] \lambda \lambda 3727, 3729$ and 
$[OIII] \lambda \lambda 4959, 5007$) contain the necessary information to  
determine accurate abundances in HII regions. In the previous study 
(Pilyugin 2000) the corresponding relations were obtained for the 
low-metallicity HII regions (12+logO/H $\leq$ 7.95, the lower branch of the 
O/H -- R$_{23}$ diagram). The high-metallicity HII regions (12+logO/H $\geq$ 
8.2, the upper branch of the O/H -- R$_{23}$ diagram) are considered in the 
present study.

A relation of the type O/H=f(P,$R_{23}$) between oxygen abundance and the 
value of abundance index R$_{23}$ introduced by Pagel 
et al. (1979) and the excitation parameter P (which is defined here as the
contribution of the radiation in $[OIII] \lambda \lambda 4959, 5007$ lines 
to the "total" oxygen radiation) has been derived empirically using the 
available oxygen abundances determined via measurement 
of a temperature-sensitive line ratio [OIII]4959,5007/[OIII]4363 ($T_{e}$ -- 
method). By comparing oxygen abundances in high-metallicity HII regions 
derived with the $T_{e}$ -- method and those derived with the suggested relations 
(P -- method), it was found that the precision of oxygen abundance 
determination with the P -- method is around 0.1 dex (the mean 
difference for the 38 HII regions considered is $\sim$ 0.08dex) 
and is comparable to that of the $T_{e}$ -- method. 

A relation of the type T$_{e}$=f(P,$R_{23}$) between electron temperature
and the values of abundance index R$_{23}$ and the excitation parameter P was 
derived empirically using the available electron temperatures 
determined via measurement of temperature-sensitive line ratios. 
The maximum value of differences between electron temperatures determined via 
measurement of temperature-sensitive line ratios and those derived with the 
suggested relation is around 1000K for HII regions considered here,
the mean value of differences for 38 HII regions is $\sim$ 500K, which is the 
same order of magnitude as the uncertainties of electron temperature determinations in 
high-metallicity HII regions via measured temperature-sensitive line ratios.


\end{abstract}

\section{Introduction}

An investigation of chemical properties of galaxies is very important for 
the development of the theory of structure and evolution of galaxies. 
Oxygen plays a key role in 
understanding the (chemical) evolution of galaxies for a several reasons. 
Firstly,  good spectrophotometry of HII regions is available now for a large 
number of galaxies, and the oxygen abundances derived from line intensities 
are published in many works
(Caplan et al. 2000;
Deharveng et al. 2000;
Esteban et al. 1998, 1999a,b;
Garnett et al. 1997; 
Izotov and Thuan 1998, 1999; 
Izotov, Thuan, and Lipovetsky 1994, 1997; 
Kobulnicky and Skillman 1996, 1997, 1998;
Kobulnicky et al. 1997; 
Skillman et al. 1994;
van Zee et al. 1997, 1998; among others).
Secondly, the origin of oxygen seems to be reliably established in contrast to
other elements like carbon or nitrogen. The oxygen abundance can be
considered as a tool to investigate the evolution of galaxies. For example, 
the value of oxygen abundance in a galaxy combined with the value of the gas 
mass fraction can tell us about the efficiency of mass
exchange between a galaxy and its environment (Pilyugin and Ferrini 1998, 2000).

Accurate oxygen abundances can be derived from  measurement of temperature-sensitive 
line ratios, such as [OIII]4959,5007/[OIII]4363. This method will be referred to as the 
T$_{e}$ - method. 
Unfortunately, in oxygen-rich HII regions the temperature-sensitive lines such as 
[OIII]4363 are too weak to be detected. For such HII regions, empirical abundance 
indicators based on more readily observable lines were suggested (Pagel et al.
1979; Alloin et al. 1979). The empirical oxygen abundance indicator 
R$_{23}$ = ([OII]3727,3729 + [OIII]4959,5007)/H$_{\beta}$, 
suggested by Pagel et al. (1979), has found widespread acceptance and use for the 
oxygen abundance determination in HII regions where the temperature-sensitive 
lines are undetectable. This method will be referred to as the R$_{23}$ - method. 
Several workers have suggested calibrations of R$_{23}$ in terms of oxygen abundance 
(Edmunds and Pagel 1984, McCall et al. 1985, Dopita and Evans 1986, 
 Zaritsky et al. 1994, among others). 

There are two problems with the oxygen abundances derived by the R$_{23}$
 -- method. First, oxygen abundances derived with different R$_{23}$ - calibrations 
can differ by 0.3dex and more. The usually-used  R$_{23}$ 
calibrations (Edmunds and Pagel 1984, McCall et al. 1985, Dopita and Evans 1986, 
McGaugh 1991) are based on a few then-available oxygen abundance determinations 
through the T$_{e}$ -- method and HII region models. More oxygen abundance 
determinations through the T$_{e}$ -- method are available now. None of these
R$_{23}$  calibrations can reproduce the available data well enough (Pilyugin 
2000, Paper I). Secondly, it has been found (Paper I) that the error in 
the oxygen abundance derived with the R$_{23}$ -- method involves two parts: 
the first is a random error and the second is a systematic error depending on 
the excitation parameter. The origin of this systematic error is as follows.
In a general case the intensities of oxygen emission lines in spectra of HII 
regions depend not only on the oxygen abundance but also on the physical 
conditions (hardness of the ionizing radiation and geometrical factor). Then in 
the determination of the oxygen abundance from line intensities the physical 
conditions in the HII region should be taken into account. In the T$_{e}$ -- method 
this is done via T$_{e}$. In the R$_{23}$ -- method the physical conditions in 
an HII region are ignored. 

In our recent work (Paper I), a new way of oxygen abundance determination 
in HII regions (P -- method) was suggested. A more general relation of 
the type O/H=f(P,R$_{23}$) is used in the P -- method, compared to the 
relation of the type O/H=f(R$_{23}$) used in the traditional R$_{23}$ -- method. 
It was found in Paper I that the precision of oxygen abundance determinations 
in low-metallicity (12+logO/H $\leq$ 7.95) HII regions with the P -- method is 
comparable to that with the T$_{e}$ -- method.
The oxygen abundances of high-metallicity (12+logO/H $\geq$ 8.2) HII regions 
derived with the P -- method are significantly less accurate. This seems to 
be because a subset of HII regions with  high-quality
homogeneous determinations of oxygen abundances was used in the construction
of the O/H=f(P,R$_{23}$) relation in the case of the low-metallicity HII regions
while the relation for high-metallicity HII regions was based on the set of all 
available HII regions with inhomogeneous determinations of oxygen abundances.
Here the oxygen abundances for high-metallicity HII regions with measured 
line ratios $[OIII] \lambda \lambda 4959, 5007 / \lambda 4363$ will be
recomputed in a uniform manner, and the set of HII regions with homogeneous 
determinations of oxygen abundances will be used in the construction of the 
O/H=f(P,R$_{23}$) relation in the case of the high-metallicity HII regions.

The search for an O/H=f(P,R$_{23}$) relation for high-metallicity HII regions 
is the goal of this study. The preliminary analysis of the relevant 
observational data and redetermination of oxygen abundance in a uniform way 
for a large set of HII regions are given in Section 2. The line intensities -- 
O/H calibration is derived in Section 3. The line intensities -- T$_{e}$ 
calibration is derived in Section 4. A discussion is presented in 
Section 5. Section 6 contains a brief summary.

\section{Preliminary analysis of observational data}

\begin{figure}[thb]
\vspace{6.5cm}
\includegraphics{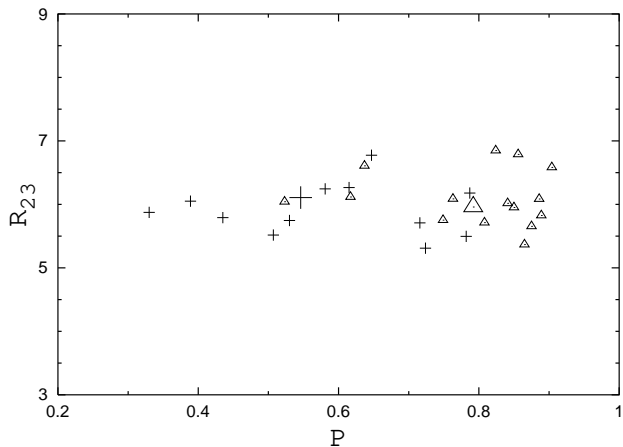}
\caption{\label{figure:10396f01}
The  P -- R$_{23}$ diagram. The small triangles are measurements for individual 
areas of the HII region M17 in the Milky Way Galaxy (Peimbert et al. 1992), the 
large triangle is the integrated data. The small pluses are measurements for 
individual areas of the HII region DEM323 in the Large Magellanic Cloud, the 
large plus is the integrated data (Oey et al. 2000).
}
\end{figure}

The strategy for construction of an empirical relation between strong line 
intensities and oxygen abundance is based on the following propositions
{\it 1)} the value of O/H in high-metallicity HII region can be expressed 
as a function of two parameters: the value of R$_{23}$ and the hardness of the 
ionizing radiation, {\it 2)} the excitation index P is a good indicator of the 
hardness of the ionizing radiation. (Notations similar to those of Paper I will 
be adopted here: $R_{2}$ = $I_{[OII] \lambda 3727+ \lambda 3729} /I_{H\beta }$, 
$R_{3}$ = $I_{[OIII] \lambda 4959+ \lambda 5007} /I_{H\beta }$, 
$R$ = $I_{[OIII] \lambda 4363} /I_{H\beta }$, $R_{23}$ =$R_{2}$ + $R_{3}$, 
$X_{23}$ = log$R_{23}$, and P = $R_{3}$/$R_{23}$. The excitation index P used 
here and indexes $p_{2}$ and $p_{3}$ used in Paper I are related through simple 
expressions: $p_{3}$ = logP and  $p_{2}$ = log(1-P).)
These propositions are immediately evident from the observational data. It has 
been shown that the value of R$_{23}$ is a robust property of a given HII 
region (Kennicutt et al. 2000, Oey et al. 2000) in the sense that its value is 
relatively constant within a given HII region. As an illustration of this fact, 
Fig.\ref{figure:10396f01} shows the R$_{23}$ -- P diagram for the multiple 
positions in the HII region M17 (Milky Way Galaxy) and in the HII region 
DEM323 (Large Magellanic Cloud). The M17 data include spectra from Peimbert et 
al. (1992) and the DEM323 data are from Oey et al. (2000). Both HII regions show 
a large range of excitation across the regions sampled; at the same time the 
value of R$_{23}$ is relatively constant. Other examples can be found in
Kennicutt et al. (2000) and Oey et al. (2000).
 
On the other hand, Fig.\ref{figure:10396f01} shows that 
the value of R$_{23}$ is not a good indicator of metal abundance in HII 
regions in the sense that there is no one-to-one correspondence
between R$_{23}$ and oxygen abundance. Indeed,
the HII regions M17 and DEM323 have in fact the same values of R$_{23}$,
while their oxygen abundances are rather different: 12+logO/H = 8.37
in DEM323 versus 12+logO/H = 8.61 in M17. Thus the value of R$_{23}$ 
is governed not only by the oxygen abundance but at least one
additional parameter. It is usually accepted that the spectrum of an HII region
(and hence the value of R$_{23}$) is specified by three parameters:
the abundance of the chemical elements in the gas, the ionization parameter
or geometrical factor, and the hardness of the ionizing radiation 
(Stasinska 1990, McGaugh 1991, among others). The relative constancy of 
R$_{23}$ within a given nebula suggests that its value in moderately 
metal-rich HII region depends very weakly (if at all) on the ionization 
parameter. This leads us to anticipate that the value of O/H is a
function of two parameters: the value of R$_{23}$ and the hardness of
the ionizing radiation. Thus, an important proposition that in moderately 
high-metallicity HII regions R$_{23}$ is mainly governed by the oxygen 
abundance and by the hardness of the ionizing radiation (or effective 
temperature of the exciting star(s)) is immediately evident from the 
following observational facts: {\it 1)} the value of R$_{23}$ is relatively 
constant within a given HII region, {\it 2)} there is no one-to-one 
correspondence between R$_{23}$ and oxygen abundance. 

\begin{figure}[thb]
\vspace{6.5cm}
\includegraphics{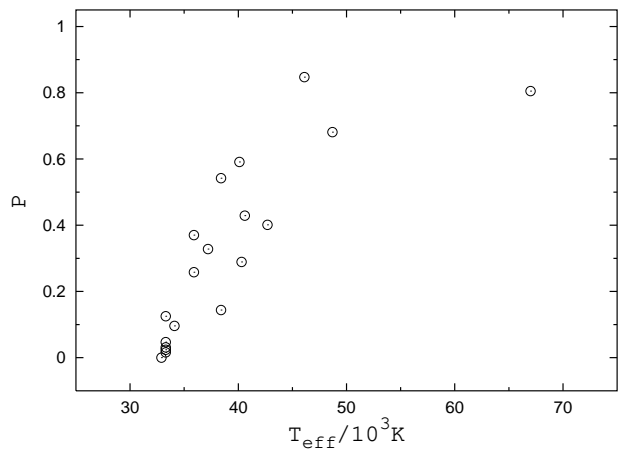}
\caption{\label{figure:10396f02}
The  P -- $T_{eff}$ diagram for the "calibrating" objects from Kennicutt et al.
(2000). 
}
\end{figure}

It has been found that the value of [OIII]/H$_{\beta}$ correlates with the 
effective temperature of the exciting star or the hardness of the ionizing 
radiation (Kaler 1978, Copetti et al. 1986, among others) and hence the value of 
{[OIII]/H$_{\beta}$ can be used as indicator of the hardness of the ionizing 
radiation. Vilchez and Pagel (1988) have introduced the value
\begin{equation}
\frac{[OII] \lambda \lambda 3727, 3729 / [OIII] \lambda \lambda 4959, 5007}
{[SII] \lambda \lambda 6716, 6731 / [SIII] \lambda \lambda 9069, 9532}
\end{equation}
as a hardness index. [OIII]/H$_{\beta}$ is not a very good indicator
of the radiation hardness because it is affected by 
the oxygen abundance as well. The hardness index of Vilchez and
Pagel contains the intensity of line [SIII]$\lambda$9532, which is often
unknown. It will be assumed here that the excitation index P
can be used as indicator of the hardness of the ionizing radiation. 
The observational evidence in favor of this proposition is given 
in Fig.\ref{figure:10396f02};  the "calibrating" objects from Kennicutt et al.
(2000) show a clearly defined correlation between the excitation index P
and effective temperature of the ionizing star. 

The spectra of part but not the whole HII region are often observed. It should 
be noted that the value  of P is a good indicator of the hardness of the 
ionizing radiation only if measured fluxes reflect their relative contributions 
to the radiation of the whole nebula. This fact, together with 
the quality of spectra, has been taken into account in compiling the sample 
of HII regions used for the construction of an empirical relation between  strong 
line intensities and oxygen abundance. If the measurements for individual 
areas in the same HII region were reported in a paper, the integrated spectra
were derived (if authors did not do so). If several spectra of equal quality 
were available for the same HII region, the spectrum with largest H$_{\beta}$ 
flux was preferred. Our sample includes 38 HII regions with 12+log(O/H)$_{T_{e}}$ 
$>$ 8.2 for which we have collected the relevant observational data, listed with 
references in Table \ref{table:data}.

\begin{table*}
\caption[]{\label{table:data}
Characteristics for the HII regions in the present sample. 
The commonly used name of the galaxy is given in column 1, the name 
of the HII region is reported in column 2. The fluxes 
$R_{2}$ = $I_{[OII] \lambda 3727+ \lambda 3729} /I_{H\beta }$, 
$R_{3}$ = $I_{[OIII] \lambda 4959+ \lambda 5007} /I_{H\beta }$, 
R = $I_{[OIII] \lambda 4363} /I_{H\beta }$ are listed in columns 3
to 5. The electron concentration N$_{e}$ is reported in column 6. 
The source for fluxes and electron concentration is listed in column 7. The 
electron temperature T$_{e}$ derived from the measured R3/R line ratio is 
given in column 8. The electron temperature t$_{P}$ determined with the 
suggested T$_{e}$=f(P,R$_{23}$) relation is reported in column 9. The electron 
temperatures are given in  units of 10$^{4}$K. The oxygen abundance 
O/H$_{T_{e}}$ derived through the T$_{e}$ -- method with T$_{e}$ is listed in 
column  10. The oxygen abundance O/H$_{P}$ determined via suggested 
O/H=f(P,R$_{23}$) relation is given in column 11. The oxygen abundance 
O/H$_{t_{P}}$ derived through the T$_{e}$ -- method with t$_{P}$ is listed 
in column  12. The oxygen abundances  are given in  units of 12+logO/H.
}
\begin{center}
\begin{tabular}{lllllrcccccc} \hline \hline
       &                    &         &         &           &        &            &         &          &               &           &                  \\  
galaxy & HII region         & R$_{2}$ & R$_{3}$ &   R       & $N_{e}$& references & $T_{e}$ & $t_{P}$  & O/H$_{T_{e}}$ & O/H$_{P}$ & O/H$_{t_{P}}$    \\   
       &                    &         &         &           &        &            &         &          &               &           &                  \\  \hline
MWG  & M16=Sh49             &   1.429 &   1.262 &   0.0012  &    124  & C+D  &  0.66 &  0.63 &  8.58 &  8.68 &  8.67 \\ 
     & Sh117                &   1.799 &   1.506 &   0.0018  &     17  & C+D  &  0.69 &  0.69 &  8.59 &  8.60 &  8.60 \\ 
     & Sh184                &   1.900 &   1.762 &   0.0032  &     65  & C+D  &  0.76 &  0.72 &  8.48 &  8.58 &  8.58 \\ 
     & Sh206                &   1.189 &   4.637 &   0.0137  &    412  & C+D  &  0.85 &  0.80 &  8.47 &  8.57 &  8.58 \\ 
     & Sh212                &   3.102 &   1.741 &   0.0068  &    126  & C+D  &  0.92 &  0.88 &  8.32 &  8.38 &  8.38 \\ 
     & M42=Orion            &   1.355 &   3.912 &   0.0101  &   3577  & C+D  &  0.82 &  0.78 &  8.55 &  8.59 &  8.64 \\ 
     & M17                  &   1.239 &   4.720 &   0.010   &    691  & P92  &  0.79 &  0.81 &  8.61 &  8.56 &  8.57 \\ 
     & S298=RCW5-1          &   0.912 &   9.119 &   0.0708  &    200  & S83  &  1.13 &  1.06 &  8.27 &  8.34 &  8.36 \\ 
     & S311=RCW16-1         &   2.344 &   3.861 &   0.0107  &    200  & S83  &  0.84 &  0.90 &  8.54 &  8.42 &  8.43 \\ 
NGC55   &  No 2             &   2.66  &   5.92  &   0.032   &    100  & W83  &  1.01 &  1.09 &  8.40 &  8.28 &  8.29 \\ 
        &  No 7             &   2.61  &   5.83  &   0.033   &    100  & W83  &  1.02 &  1.07 &  8.37 &  8.29 &  8.30 \\ 
NGC300  &  No 15            &   2.90  &   5.56  &   0.044   &    100  & W83  &  1.14 &  1.10 &  8.23 &  8.26 &  8.28 \\ 
NGC598  &  NGC588           &   1.482 &   6.310 &   0.024   &    100  & V88  &  0.91 &  0.94 &  8.48 &  8.43 &  8.44 \\ 
        &  NGC604           &   2.152 &   2.852 &   0.0075  &    100  & D87  &  0.83 &  0.81 &  8.47 &  8.50 &  8.50 \\ 
NGC925  & -022+227          &   2.04  &   5.653 &   0.022   &    100  & vZ98 &  0.92 &  0.98 &  8.48 &  8.38 &  8.39 \\ 
NGC1313 &  No 5             &   1.66  &   6.26  &   0.026   &    100  & P80  &  0.93 &  0.96 &  8.46 &  8.40 &  8.41 \\ 
        &  No 7             &   3.63  &   4.70  &   0.039   &    100  & P80  &  1.16 &  1.17 &  8.22 &  8.20 &  8.21 \\ 
NGC1569 &  C6w              &   1.112 &   9.018 &   0.077   &    100  & K97  &  1.17 &  1.08 &  8.23 &  8.31 &  8.34 \\ 
NGC2403 &  VS 35            &   2.46  &   1.84  &   0.0051  &    100  & G97  &  0.84 &  0.79 &  8.40 &  8.48 &  8.48 \\ 
        &  VS 38            &   1.92  &   1.34  &   0.0025  &    100  & G97  &  0.76 &  0.69 &  8.43 &  8.59 &  8.59 \\ 
        &  VS 44            &   2.8   &   1.99  &   0.0063  &    100  & G97  &  0.87 &  0.85 &  8.39 &  8.42 &  8.42 \\ 
        &  VS 51            &   2.28  &   2.37  &   0.0054  &    100  & G97  &  0.80 &  0.80 &  8.50 &  8.50 &  8.50 \\ 
        &  VS 3             &   2.26  &   2.14  &   0.0068  &    100  & G97  &  0.87 &  0.79 &  8.35 &  8.51 &  8.51 \\ 
NGC2805 & +037-115          &   2.52  &   4.662 &   0.020   &    100  & vZ98 &  0.94 &  0.98 &  8.42 &  8.36 &  8.37 \\ 
        & -068-079          &   3.64  &   4.552 &   0.033   &    100  & vZ98 &  1.11 &  1.16 &  8.26 &  8.21 &  8.22 \\ 
        & +116-098          &   3.11  &   4.681 &   0.030   &    100  & vZ98 &  1.07 &  1.07 &  8.29 &  8.27 &  8.28 \\ 
NGC4214 & Knot2 (A6)        &   3.172 &   3.746 &   0.021   &    100  & K96  &  1.02 &  1.02 &  8.31 &  8.31 &  8.31 \\ 
        & Knot5 (C6)        &   2.557 &   5.371 &   0.024   &    100  & K96  &  0.96 &  1.03 &  8.44 &  8.32 &  8.33 \\ 
NGC4395 & -272+186          &   3.06  &   4.968 &   0.024   &    100  & vZ98 &  0.98 &  1.09 &  8.42 &  8.27 &  8.28 \\ 
NGC5457 &  NGC5447          &   1.90  &   5.50  &   0.028   &    100  & S75  &  0.99 &  0.95 &  8.35 &  8.41 &  8.42 \\ 
        &  NGC5455          &   3.09  &   5.25  &   0.028   &    100  & S75  &  1.01 &  1.11 &  8.40 &  8.25 &  8.26 \\ 
        &  NGC5461          &   2.14  &   4.04  &   0.015   &    234  & T89  &  0.90 &  0.89 &  8.42 &  8.45 &  8.45 \\ 
        & -347+276          &   0.98  &   7.906 &   0.036   &    100  & vZ98 &  0.96 &  0.98 &  8.45 &  8.40 &  8.42 \\ 
        & -459-053          &   3.00  &   3.996 &   0.022   &    100  & vZ98 &  1.02 &  1.01 &  8.32 &  8.32 &  8.32 \\ 
LMC     & DEM323            &   2.770 &   3.337 &   0.014   &     10  & O+O  &  0.94 &  0.93 &  8.37 &  8.38 &  8.38 \\ 
UM311   &                   &   1.800 &   5.295 &   0.023   &    100  & I98  &  0.95 &  0.92 &  8.40 &  8.44 &  8.44 \\ 
UGC2984 & No 2              &   2.592 &   6.499 &   0.048   &    100  & vZ97 &  1.12 &  1.12 &  8.28 &  8.26 &  8.28 \\ 
UGC5716 & No 1              &   2.640 &   5.025 &   0.031   &    100  & vZ97 &  1.05 &  1.02 &  8.30 &  8.32 &  8.33 \\  \hline
\end{tabular}
\end{center}

\vspace{0.1cm}

{\it List of references }:

C+D  -- Caplan et al. (2000) + Deharveng et al. (2000);
D87  -- Diaz et al. (1987);
G97  -- Garnett et al. (1997);
I98  -- Izotov and Thuan (1998);
K96  -- Kobulnicky and Skillman (1996);
K97  -- Kobulnicky and Skillman (1997);
O+O  -- Oey and Shields (2000) + Oey et al. (2000);
P80  -- Pagel et al. (1980);
P92  -- Peimbert et al. (1992);
S75  -- Smith (1975);
S83  -- Shaver et al. (1983);
T89  -- Torres-Peimbert et al. (1989);
V88  -- Vilchez et al. (1988);
vZ97 -- van Zee et al. (1997);
vZ98 -- van Zee et al. (1998);
W83  -- Webster and Smith (1983).

\end{table*}

It is widely accepted that an accurate oxygen abundance can be derived from 
measurement of temperature-sensitive line ratios, such as   
$[OIII] \lambda \lambda 4959, 5007 / \lambda 4363$, i.e through the T$_{e}$
-- method. In the general case, the oxygen abundances in the same HII region 
with measured line ratios $[OIII] \lambda \lambda 4959, 5007 / \lambda 4363$ 
derived in different works can differ for three reasons: atomic data adopted, 
interpretation of the temperature structure (single characteristic $T_{e}$, 
two-zone model for $T_{e}$, model with small-scale temperature fluctuations) and 
errors in the line intensity measurements. Therefore the compilation of HII 
regions with original oxygen abundance determinations through the T$_{e}$ -- 
method from different works carried out over more than twenty years is not a set
of homogeneous determinations. Accordingly, 
the available published spectra of HII regions with measured line ratios 
$[OIII] \lambda \lambda 4959, 5007 / \lambda 4363$ (listed in the 
Table \ref{table:data}) have been reanalysed to produce a homogeneous set.
Two-zone models of HII regions with the algorithm
for oxygen abundance determination from Pagel et al. (1992) 
and T$_{e}$([OII]) -- T$_{e}$([OIII]) relation from Garnett (1992)
were adopted here. The oxygen abundances for HII regions from
Table \ref{table:data}  were  recomputed in this common way. 
The obtained electron temperature $T_{e}$ and corresponding oxygen abundance 
O/H$_{T_{e}}$ are reported in columns 8 and 10 of Table \ref{table:data}.
The electron temperatures derived by Garnett et al. (1997) were used
for NGC2403 HII regions (the R = $I_{[OIII] \lambda 4363} /I_{H\beta }$ values
for NGC2403 HII regions reported in Table \ref{table:data} correspond to
these electron temperatures but not to the measurements). This set of HII 
regions with homogeneous determinations will be used in the construction of an 
empirical relation between  strong line intensities and oxygen abundance.

\section{Line intensities -- O/H calibration}

The $X_{23}$ versus O/H diagram for HII regions from Table \ref{table:data} is 
presented in Fig.\ref{figure:10396f03}. The open circles are HII regions with  
P $>$ 0.5, the filled circles are HII regions with  P $<$ 0.5. The line 
is the R$_{23}$ calibration of Edmunds and Pagel (1984). Inspection of 
Fig.\ref{figure:10396f03} shows that there is no one-to-one correspondence
between $X_{23}$ value and oxygen abundance. For a fixed oxygen abundance
the positions of low-excitation HII regions in the $X_{23}$ -- O/H diagram are 
systematically shifted towards lower values of $X_{23}$ compared to the
positions of high-excitation HII regions. The calibration of Edmunds and Pagel 
(1984) corresponds to the positions of high-excitation HII regions. This is not
suprising since observational data only for high-excitation HII regions were
available when this calibration was suggested.
Other previous calibrations (McCall \& Rybski and Shields 1985, 
Dopita \& Evans 1986, Zaritsky et al. 1994) are shifted towards still
higher oxygen abundances. The fact that for a fixed oxygen abundance the value 
of $X_{23}$ varies with the excitation parameter P confirms our proposition
that the excitation index P can be used in the oxygen abundance determination.

\begin{figure}[thb]
\vspace{6.5cm}
\includegraphics{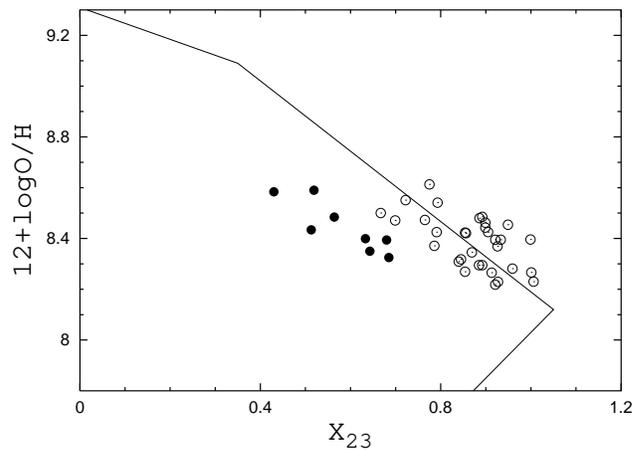}
\caption{\label{figure:10396f03}
The  X$_{23}$ -- O/H diagram for the HII regions from our sample. The filled 
circles are HII regions with P $<$ 0.5, open circles are HII regions with  
P $>$ 0.5. The line is the R$_{23}$ calibration after Edmunds and Pagel (1984).
}
\end{figure}

\begin{figure}[thb]
\vspace{9.0cm}
\includegraphics{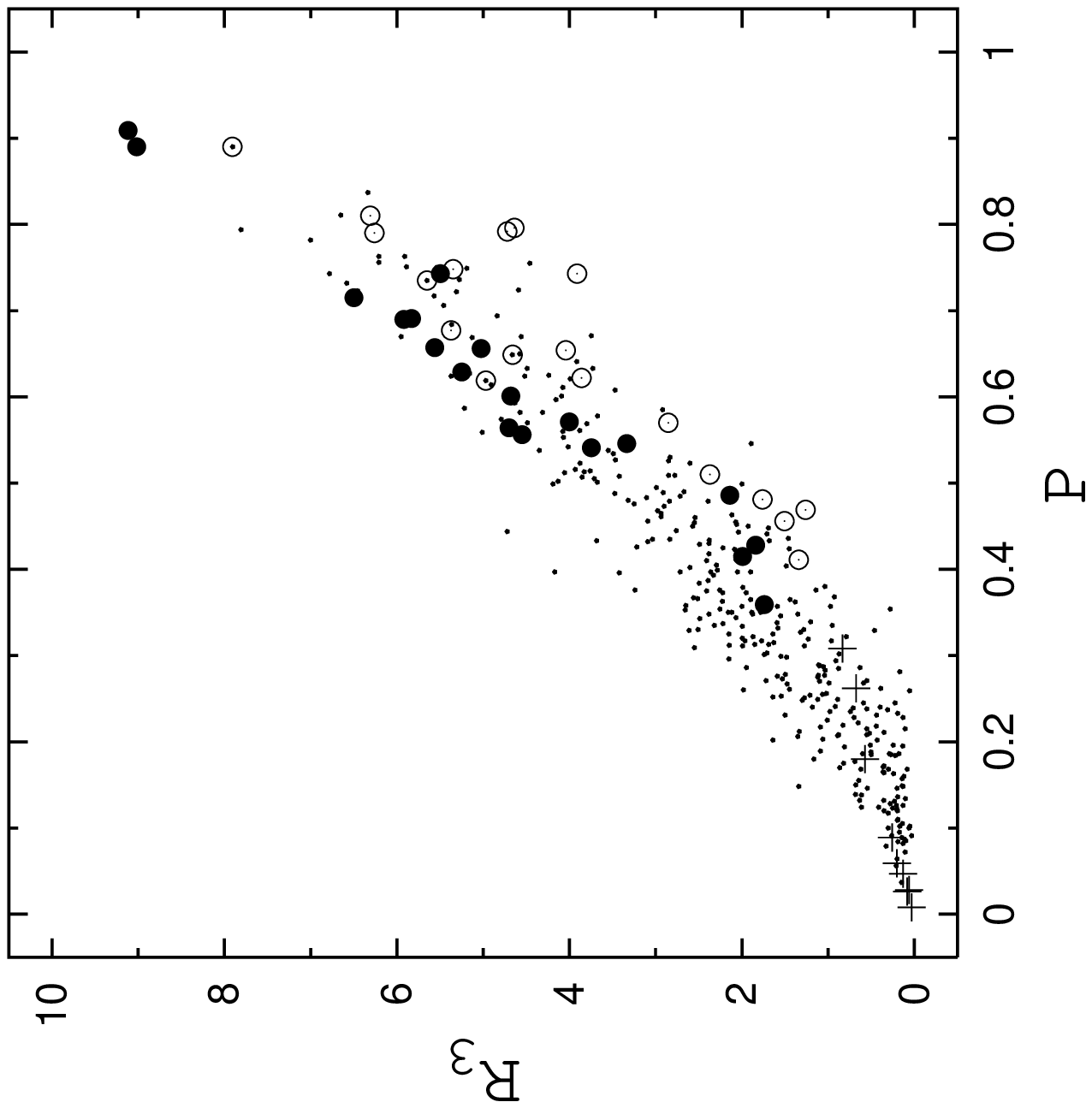}
\caption{\label{figure:10396f04}
The  P - R$_{3}$ diagram. The HII regions from our sample are represented by 
circles; the filled circles are HII regions with  12+log(O/H) $<$ 8.4, open 
circles are HII regions with  12+log(O/H) $>$ 8.4. The points are HII regions 
from Zaritsky et al. (1994) and van Zee et al. (1998). The pluses are low-excitation 
HII regions in our Galaxy from Caplan et al. (2000), Deharveng et al. (2000).
}
\end{figure}

\begin{figure}[thb]
\vspace{9.0cm}
\includegraphics{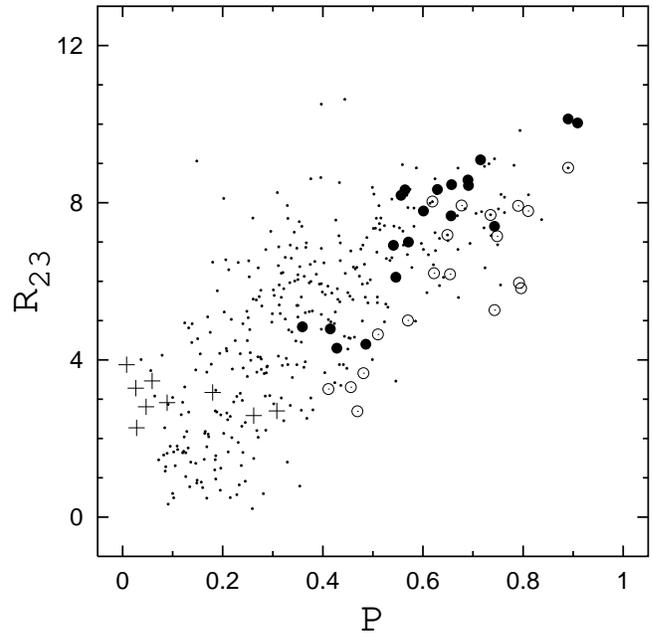}
\caption{\label{figure:10396f05}
The  P - R$_{23}$ diagram. The HII regions from our sample are represented by 
circles; the filled circles are HII regions with  12+log(O/H) $<$ 8.4, open 
circles are HII regions with  12+log(O/H) $>$ 8.4. The points are HII regions 
from Zaritsky et al. (1994) and van Zee et al. (1998). The pluses are low-excitation 
HII regions in our Galaxy from Caplan et al. (2000), Deharveng et al. (2000).
}
\end{figure}

It is convenient to start the search for the relation between oxygen abundance
and parameters P and R$_{23}$ from consideration of R$_{3}$ -- P
(Fig.\ref{figure:10396f04}) and R$_{23}$ -- P (Fig.\ref{figure:10396f05}) 
diagrams. The positions of HII regions from Table \ref{table:data} in these diagrams are 
presented by circles; the filled circles are HII regions with 
12+log(O/H) $<$ 8.4, open circles are HII regions with 12+log(O/H) 
$>$ 8.4. The points are HII regions from Zaritsky et al. (1994) and van Zee et al.
(1998). The pluses are low-excitation HII regions in our Galaxy from
Deharveng et al. (2000). Fig.\ref{figure:10396f04} shows that HII regions occupy
rather narrow band in the R$_{3}$ -- P diagram. The relation between R$_{3}$ and 
P can be given by a polynomial
\begin{equation}
k_{0}R_{3} = k_{1} P + k_{2} P^{2} + k_{3} P^{3}.
\label{equation:kR3}
\end{equation}
The zero-degree term of the polynomial must be equal to zero since R$_{3}$ and
P are equal to zero simultaneously by the definition. 
Examination of Fig.\ref{figure:10396f04} shows that the positions of 
the HII regions with  12+log(O/H) $<$ 8.4 are shifted relative to those 
with  12+log(O/H) $>$ 8.4. In order to take this fact into account the coefficients 
of Eq.(\ref{equation:kR3}) will be taken in the form
\begin{equation}
k_{j} = a_{j}  + b_{j} Z ,
\label{equation:k}
\end{equation}
where the notation Z=12+logO/H is used for brevity. Taking Eq.(\ref{equation:k}) 
into account, Eq.(\ref{equation:kR3}) can be rewritten as 
\begin{equation}
R_{3} = \frac{(a_{1} +b_{1} Z) P + (a_{2} +b_{2} Z) P^{2} + 
(a_{3} +b_{3} Z) P^{3}}{1 + b_{0} Z}.
\label{equation:R3}
\end{equation}
The coefficient a$_{0}$ has been taken equal to 1 (deviding the numerator and
denominator in the right side of Eq.(\ref{equation:R3}) by a$_{0}$). 
Eq.(\ref{equation:R3}) can be solved for the value Z=12+logO/H 
\begin{equation}
12+log(O/H)_{P} = \frac{R_{3} - a_{1} P -a_{2} P^{2} - a_{3} P^{3}}
                       {b_{1} P +b_{2} P^{2} + b_{3} P^{3}-b_{0} R_{3}}  .
\label{equation:OHR3}
\end{equation}
Taking into account that $R_{3} = P \times R_{23}$, Eq.(\ref{equation:OHR3}) 
can be transformed into 
\begin{equation}
12+log(O/H)_{P} = \frac{R_{23} - a_{1}  -a_{2} P - a_{3} P^{2}}
                       {b_{1}  +b_{2} P + b_{3} P^{2}-b_{0} R_{23}}  .
\label{equation:OHR23}
\end{equation}
The coefficients $b_{0}$, $a_{1}$, $b_{1}$, $a_{2}$, $b_{2}$, $a_{3}$, $b_{3}$
can be found using the set of HII regions with oxygen abundances derived through
the T$_{e}$ -- method. In other words, the positions in the R$_{23}$ -- P
(and the R$_{3}$ -- P) diagram can be calibrated in terms of oxygen abundance.

\begin{figure}[thb]
\vspace{6.5cm}
\includegraphics{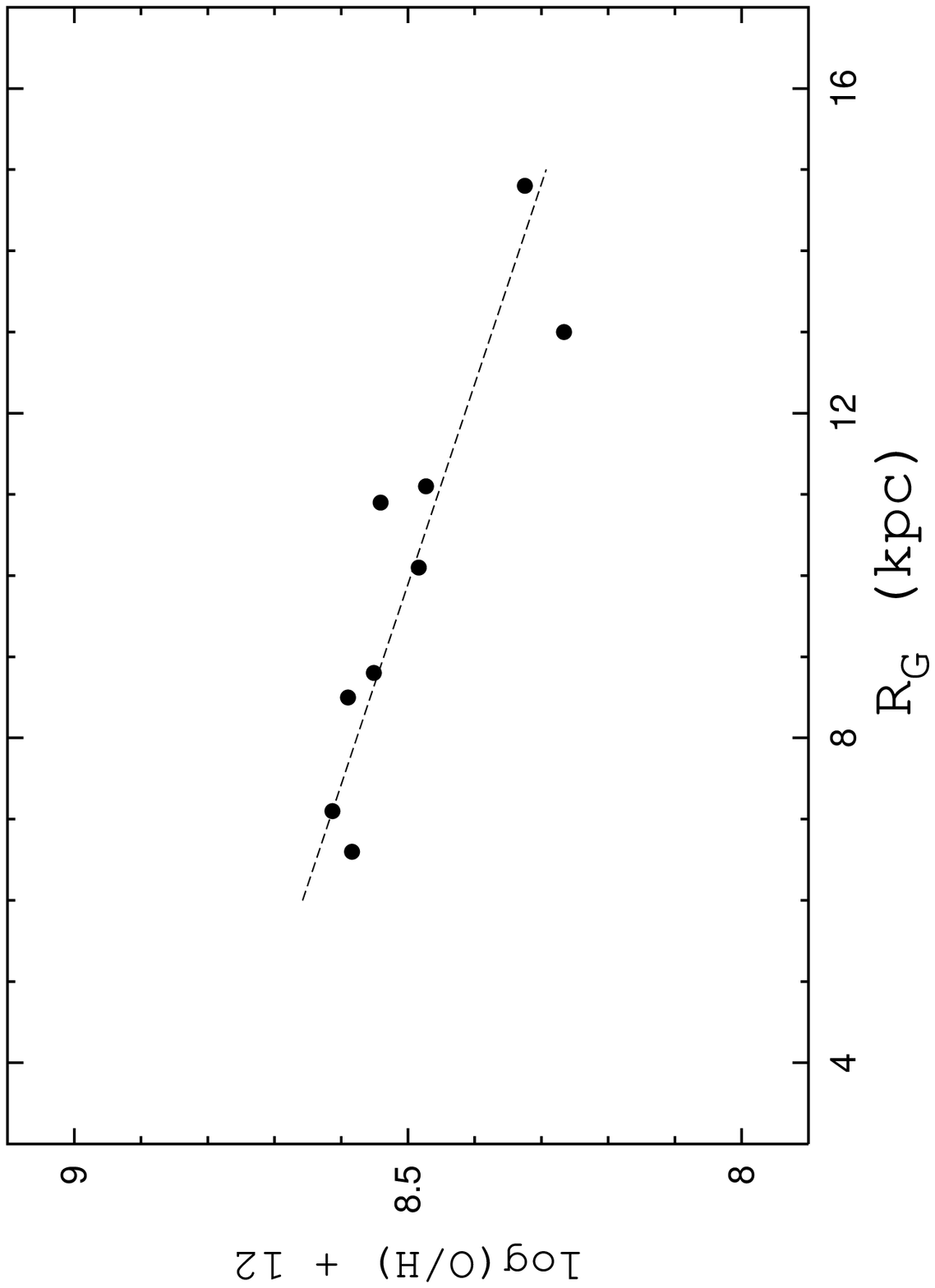}
\caption{\label{figure:10396f06}
The radial distribution of oxygen abundance in the disk of our Galaxy. The 
oxygen abundances of HII regions are determined through the
T$_{e}$ -- method (recomputed here); the distances are taken from Deharveng
et al. (2000). 
}
\end{figure}

\begin{figure}[thb]
\vspace{10.2cm}
\includegraphics{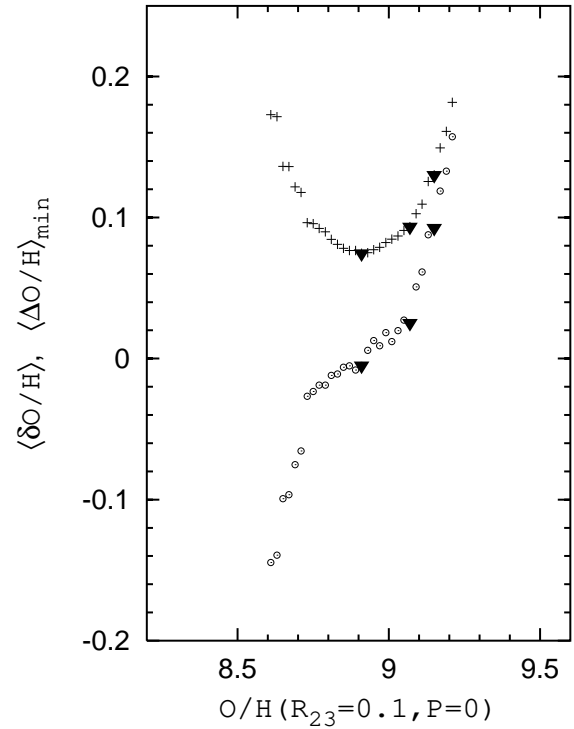}
\caption{\label{figure:10396f07}
The $\left\langle \Delta O/H \right\rangle _{min}$ and 
$\left\langle \delta O/H \right\rangle$ as a function of the 
O/H$^{*}$ = O/H(R$_{23}$=0.1,P=0). 
The $\left\langle \Delta O/H \right\rangle _{min}$ values are presented by 
pluses, the $\left\langle \delta O/H \right\rangle$ values are presented by 
the circles. The variants of the calibration (V891, V907, and V915) discussed in 
the text are shown by filled triangles.
}
\end{figure}

The precision of present-day determinations of the oxygen abundances
in high-metallicity HII regions through the T$_{e}$ -- method seems to
be around 0.1dex (Deharveng et al. 2000). It has been found that there are
different sets of coefficients $b_{0}$, $a_{1}$, $b_{1}$, $a_{2}$, $b_{2}$, 
$a_{3}$, $b_{3}$ (or different variants of the calibration)
which give an average value of differences
$\Delta$O/H$_{P}$ = logO/H$_{P}$ -- logO/H$_{T_{e}}$ less than 0.1 dex
for our sample of HII regions. Unfortunately, our sample of HII regions (used 
in the search for the coefficients) does not contain the HII regions with 
P $<$ 0.4, Fig.\ref{figure:10396f05}. Therefore different sets of 
coefficients based on the narrow range of oxygen abundances in high-excitation 
HII regions can result in appreciably different abundances in low-excitation 
ones although all the variants of the calibration that resulted in a correlation 
between the differences $\Delta$O/H$_{j}$ for
individual HII regions and P$_{j}$ or between $\Delta$O/H$_{j}$ and
O/H$_{j}$ were rejected and only the variants in which
both correlation coefficients are less than 0.1 were considered.
Thus, the problem of choice of the variant of the calibration, which results in
correct oxygen abundances in the whole range of P, appears. This 
difficulty is usually resolved by adding HII region models to the real HII 
regions. As was discussed by Stasinska (2000), the existing models of HII 
regions may be far from reality for a number of reasons. Then we may have to try 
to overcome this difficulty starting from the observational data only. The 
following solution to this problem will be adopted. Our sample contains
9 HII regions of the disk of our Galaxy. Fig.\ref{figure:10396f06} shows the 
oxygen abundance as a function of the galactocentric distance. The linear fit
\begin{equation}
12+log(O/H) = 8.90 - 0.041 R_{G}
\label{equation:grad}
\end{equation}
is close to the relation obtained by Deharveng et al.(2000). This is not 
surprising, since the measured fluxes for 6 out of the 9 HII regions were 
taken from them. The list of HII regions observed by Caplan et al. (2000) and 
Deharveng et al. (2000) contains a number of low-excitation ones. Following
Deharveng et al. (2000) all the heavily reddened HII regions were excluded from
consideration. The nine low-excitation HII regions Sh54, Sh131, Sh148, Sh 152,
Sh153, Sh156, Sh168, Sh217, and Sh219 for which flux measurements seems to
be reliable were selected and used as  "secondary calibrating objects" in the 
choice of the relation between strong line intensities and oxygen abundance 
(the positions of these HII regions in the R$_{3}$ -- P and R$_{23}$ -- P 
diagrams are shown by pluses, Figs.\ref{figure:10396f04}, \ref{figure:10396f05}). 
The basic requirement imposed on the calibration by our set of HII regions
is complemented by the condition that the oxygen abundances of selected 
low-excitation HII regions derived through the calibration must result in the 
same radial oxygen abundance gradient as was determined from HII regions with 
oxygen abundances derived through the T$_{e}$ -- method. Inspection of  
Fig.\ref{figure:10396f05} shows that positions of HII regions from this 
"extended" set cover the whole range of P. 

\begin{figure*}
\vspace{19.2cm}
\includegraphics{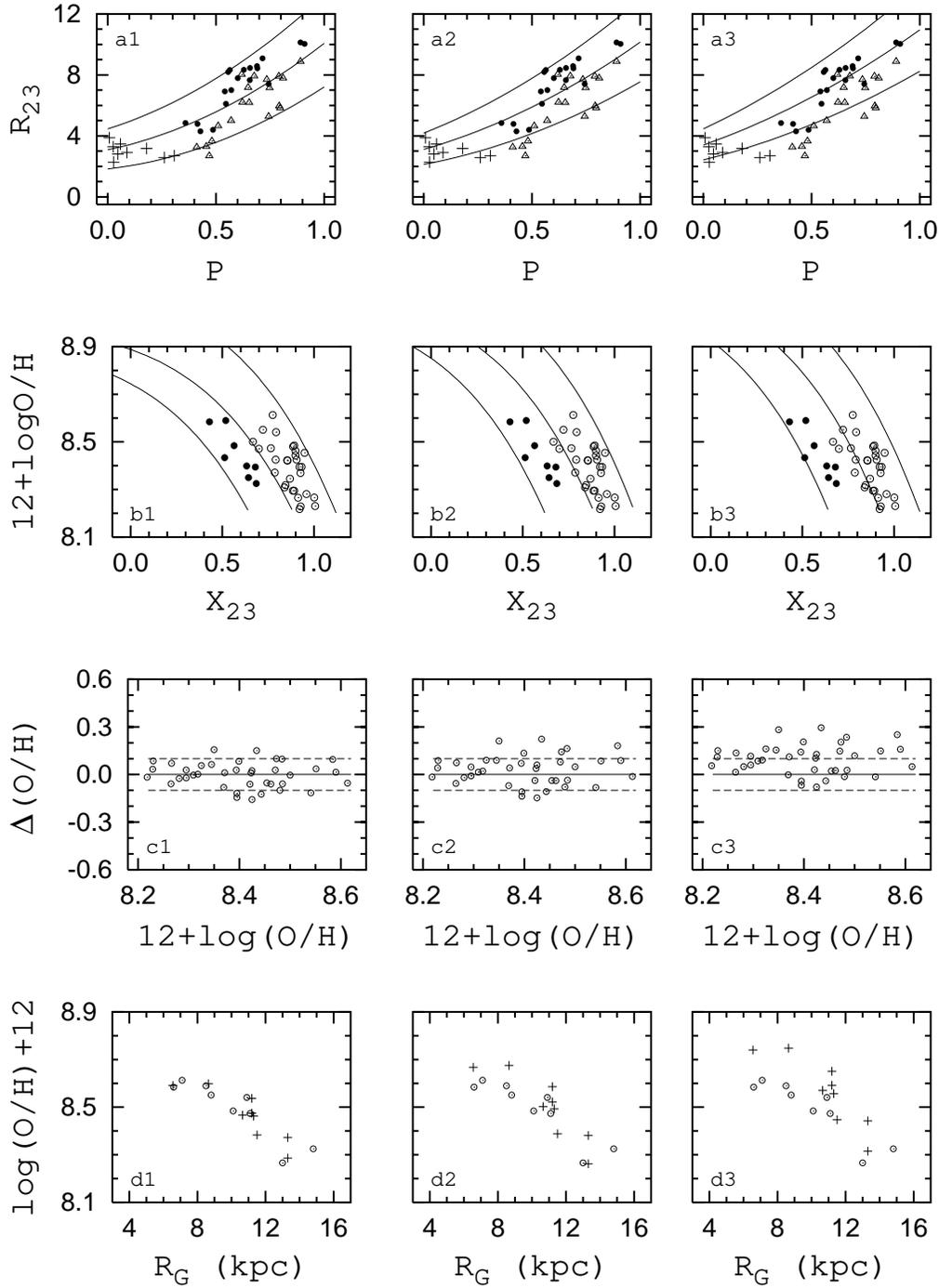}
\caption{\label{figure:10396f08}
The comparison of three variants of the calibration (V891 - 1st column panels,
V907 - 2nd column panels, V915 - 3rd column panels) with observational data. 
The panels a1, a2, and a3 show the R$_{23}$- P diagrams. The filled circles 
are HII regions with 8.2 $<$ 12+logO/H $<$ 8.4, the open triangles are HII 
regions with 8.4 $<$ 12+logO/H $<$ 8.62 from our sample. The pluses are 
low-excitation HII regions in the disk of our Galaxy from Deharveng et al.(2000).
The curves are R$_{23}$ -- P relations predicted by a given variant of the 
calibration for fixed values of logO/H+12 = 8.2, logO/H+12 = 8.4, 
logO/H+12 = 8.6 (from top to bottom).
The panels b1, b2, and b3 show the  X$_{23}$ -- O/H diagrams. The filled 
circles are HII regions with P $<$ 0.5, the open circles are HII regions with  
P $>$ 0.5 from our sample. The curves are X$_{23}$ - O/H relations predicted 
by a given variant of the calibration for 
fixed values of P=1, P=0.5, and P=0 (from top to bottom).
The panels c1, c2, and c3 show $\Delta$O/H = log(O/H)$_{P}$ -
log(O/H)$_{T_{e}}$ versus O/H$_{T_{e}}$ diagrams for our sample of HII regions.
The panels d1, d2, and d3 show the radial distribution of oxygen abundance
within the disk of our Galaxy. The circles are HII regions with oxygen
abundances derived through the $T_{e}$ -- method, the pluses are low-excitation
HII regions with oxygen abundances determined through the given variant of the 
present calibration.
}
\end{figure*}

The choice of the variant of the calibration which is suitable for the whole 
range of the parameter P was performed in the following way. A variant of the 
calibration can be characterized by three parameters; O/H$^{*}$ = 
O/H(R$_{23}$=0.1,P=0), $\left\langle \Delta O/H \right\rangle$  =  
$\sqrt{(\sum\limits_{j=1}^n (\Delta O/H_j)^2)/n}$, and
$\left\langle \delta O/H \right\rangle$  =  
$(\sum\limits_{j=1}^n \Delta O/H_j)/n$
where $\Delta O/H_j$ is equal to  logO/H$^{j}_{P}$ -- logO/H$^{j}_{T_{e}}$
for our set of HII regions and $\Delta O/H_j$ is equal to logO/H$^{j}_{P}$ -- 
logO/H$^{j}_{R_{G}}$ for the selected low-excitation HII regions where
O/H$^{j}_{R_{G}}$ is the abundance estimated through the 
galactocentric distance (Eq.(\ref{equation:grad})). The O/H$^{*}$ is the minimum 
value of oxygen abundance for HII regions with R$_{23}$=0.1, predicted by a 
given variant of the calibration. 
The value of $\left\langle \Delta O/H \right\rangle$ characterizes the scatter
of differences between oxygen abundances determined via a fixed variant 
of the calibration and adopted oxygen abundances. 
The value of $\left\langle \delta O/H \right\rangle$ characterizes the average
systematic difference between oxygen abundances determined via the fixed variant 
of the calibration and adopted oxygen abundances. For every fixed value of the 
parameter O/H$^{*}$ the variant of the calibration which gives the minimum value 
of $\left\langle \Delta O/H \right\rangle$ was obtained.
The value of $\left\langle \Delta O/H \right\rangle _{min}$ and corresponding
value of $\left\langle \delta O/H \right\rangle$ as a function of the O/H$^{*}$ 
are presented in Fig.\ref{figure:10396f07}. The local non-smooth variations of 
the values of $\left\langle \Delta O/H \right\rangle _{min}$ and 
$\left\langle \delta O/H \right\rangle$ with O/H$^{*}$ 
reflect the fact that all the variants of the calibration which
result in a correlation between $\Delta$O/H$_{j}$ and P$_{j}$ or between 
$\Delta$O/H$_{j}$ and O/H$_{j}$ were rejected and only the variants in which
both correlation coefficients are less than 0.1 were considered.

Fig.\ref{figure:10396f07} shows that the 
$\left\langle \Delta O/H \right\rangle _{min}$ has a minimum value 
for O/H$^{*}$ = 8.91. The corresponding value of 
$\left\langle \delta O/H \right\rangle$ is close to zero ($\approx$ --0.005dex).
This variant of the calibration will be referred to as V891. The comparison of 
this variant of the calibration with observational data is given in 
Fig.\ref{figure:10396f08} (1st column). Two other variants V907 and V915 of the 
calibration are also presented in Fig.\ref{figure:10396f08} (panels in columns 2 
and 3).  Fig.\ref{figure:10396f08} shows that 
the V891 variant of the calibration well reproduces the observational data of 
different types. The positions of HII galaxies from our sample in the
R$_{23}$ -- P and O/H -- R$_{23}$ diagrams are well reproduced by the 
V891 variant of the calibration (panels a1 and b1 in Fig.\ref{figure:10396f08}). 
The differences between oxygen abundances determined through the 
variant V891 of the calibration and through the T$_{e}$ 
-- method do not exceed 0.1dex (the average difference is about 0.08 dex 
for 38 HII regions) for the majority of our set (panel c1 in 
Fig.\ref{figure:10396f08}). For comparison, the average difference between 
variant V915 and the T$_{e}$ -- method is in excess of 0.1dex 
($\approx$ 0.13dex) (panel c3 in Fig.\ref{figure:10396f08}). Consequently, 
variant V891 of the calibration results in the radial
oxygen abundance gradient which is very close to that derived from HII
regions with oxygen abundances determined through the T$_{e}$ -- method
(panel d1 in Fig.\ref{figure:10396f08}). 
Then, the variant V891 of the calibration is the most credible relation
between strong line intensities and oxygen abundance.
This variant of the calibration with coefficients 
b$_{0}$ = --0.243, a$_{1}$ = --54.2, b$_{1}$ = 6.07, a$_{2}$ = --59.45,
b$_{2}$ = 6.71, a$_{3}$ = --7.31, b$_{3}$ = 0.371 and corresponding
equation
\begin{equation}
12+log(O/H)_{P} = \frac{R_{23} + 54.2  + 59.45 P + 7.31 P^{2}}
                       {6.07  + 6.71 P + 0.371 P^{2} + 0.243 R_{23}}
\label{equation:OHR23f}
\end{equation}
can be adopted for oxygen abundance determinations in moderately high-metallicity
HII regions with undetectable temperature-sensitive line ratios.
The oxygen abundances in HII regions from our sample determined with this 
calibration are given in Table \ref{table:data} (column 11). 

\section{Line intensities -- T$_{e}$ calibration}

The oxygen abundances O/H$_{P}$ derived in the previous section through the 
excitation parameter P and abundance index R$_{23}$ are in agreement with those 
derived through the classical Te -- method. This justifies the use of relations 
of the type O/H=f(P,R$_{23}$) for oxygen abundance determination in high-metallicity
HII regions with the undetectable [OIII]4363 line. It also provides evidence
that the propositions which are at the basis of this relation
({\it i)}  that the value of abundance index is mainly governed by the 
oxygen abundance and by the hardness of the ionizing radiation and
depends very weakly (if at all) on the ionization parameter, 
{\it ii)}  that the parameter P can be used as indicator
of the hardness of the ionizing radiation) seem to be close to reality.
Conversely, if the physical conditions in high-metallicity nebulae are governed 
mainly by the oxygen abundance and by the hardness of the ionizing radiation and 
if observational values of P and R$_{23}$ reflect these two, then it can be 
expected that the physical conditions in a nebula can be derived with help of
observational values P and R$_{23}$.  The physical conditions in a nebula
are reflected in the electron temperature, so that a relation of the type
Te=f(P,R$_{23}$) can be expected.

\begin{figure}[thb]
\vspace{8.8cm}
\includegraphics{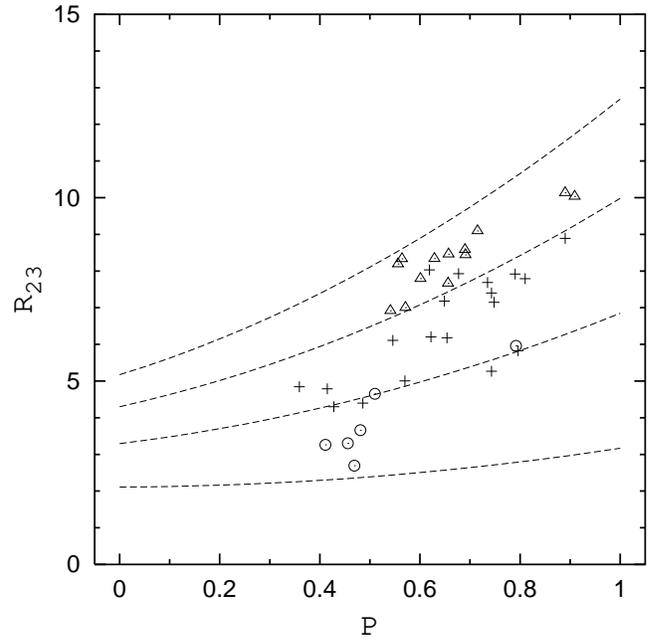}
\caption{\label{figure:10396f09}
The  P - R$_{23}$ diagram. 
The HII regions with T$_{e}$ $<$ 0.8 are presented by circles,
those with 0.8 $<$ T$_{e}$ $<$ 1.0 are shown by pluses and 
those with T$_{e}$ $>$ 1.0 are shown by triangles.
The curves are the R$_{23}$ -- P relations predicted by the present calibration 
for a fixed values of T$_{e}$ = 0.6, 0.8, 1.0, and 1.2 (from bottom to top).
}
\end{figure}

Fig.\ref{figure:10396f09} shows the positions of HII regions from 
Table \ref{table:data} in the R$_{23}$ -- P diagram. The HII regions with T$_{e}$ 
$<$ 0.8 are presented by circles, those with 0.8 $<$ T$_{e}$ $<$ 1.0 are shown 
by pluses and those with T$_{e}$ $>$ 1.0 are shown by triangles. 
Fig.\ref{figure:10396f09} shows that the HII regions with different electron 
temperatures occupy different bands in the R$_{23}$ -- P diagram. This suggests 
that the positions in  the R$_{23}$ -- P diagram can be calibrated in terms of 
electron temperatures. Let us again start from the expression of the type
\begin{equation}
R_{3} = \frac{(A_{1} +B_{1} t) P + (A_{2} +B_{2} t) P^{2} + 
(A_{3} +B_{3} t) P^{3}}{1 + B_{0} t},
\label{equation:R3t}
\end{equation}
which can be solved for the value of t
\begin{equation}
t_{P} = \frac{R_{3} - A_{1} P -A_{2} P^{2} - A_{3} P^{3}}
         {B_{1} P +B_{2} P^{2} + B_{3} P^{3}-B_{0} R_{3}},
\label{equation:t}
\end{equation}
where t$_{P}$ is the electron temperature expressed in units of 10$^{4}$K.
The coefficients B$_{0}$, A$_{1}$, B$_{1}$, A$_{2}$, B$_{2}$, A$_{3}$, and B$_{3}$
can be found using the sample of HII regions with electron temperature derived 
via measured temperature-sensitive line ratios [OIII]4959,5007/[OIII]4363. 
It has been found that there are different sets of coefficients B$_{0}$, 
A$_{1}$, B$_{1}$, A$_{2}$, B$_{2}$, A$_{3}$, and B$_{3}$ which provide the 
average value of differences $\Delta$t$_{P}$ = t$_{P}$ -- T$_{e}$ around
500K. The variant of the calibration (the choice is explained below) with 
coefficients B$_{0}$ = 0.583, A$_{1}$ = --3.09, B$_{1}$ = 9.90, A$_{2}$ = --7.05, 
B$_{2}$ = 11.86, A$_{3}$ = --2.87, B$_{3}$ = 7.05 and corresponding equation
\begin{equation}
t_{P} = \frac{R_{23} + 3.09  + 7.05 P + 2.87 P^{2}}
                       {9.90  + 11.86 P + 7.05 P^{2} - 0.583 R_{23}}
\label{equation:tf}
\end{equation}
has been adopted for the electron temperature determination in high-metallicity 
HII regions with undetectable temperature-sensitive line ratios.
The electron temperatures t$_{P}$ in HII regions from our sample determined from
Eq.(\ref{equation:tf}) are given in Table \ref{table:data} (column 9). 

\begin{figure}[thb]
\vspace{6.5cm}
\includegraphics{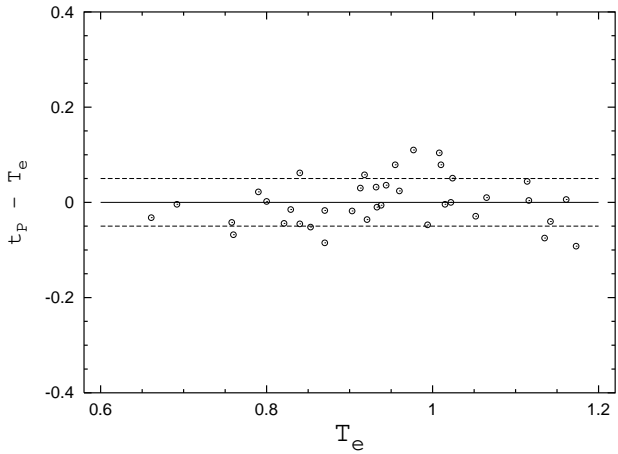}
\caption{\label{figure:10396f10}
The differences t$_{P}$ -- T$_{e}$  as a function of T$_{e}$, for our set of HII regions. 
}
\end{figure}

The R$_{23}$ -- P relations predicted by the calibration for fixed values of
t$_{P}$ = 0.6, 0.8, 1.0, and 1.2 are presented in 
Fig.\ref{figure:10396f09} by dashed lines. 
Fig.\ref{figure:10396f10} shows the differences between electron temperatures
t$_{P}$ derived through the present calibration and measured electron 
temperatures T$_{e}$. As can be seen in Fig.\ref{figure:10396f10}, the 
largest value of the difference $\Delta T_{e}$ = t$_{P}$ -- T$_{e}$  is around 
1000K, the average value about 500K.

\begin{figure}[thb]
\vspace{6.5cm}
\includegraphics{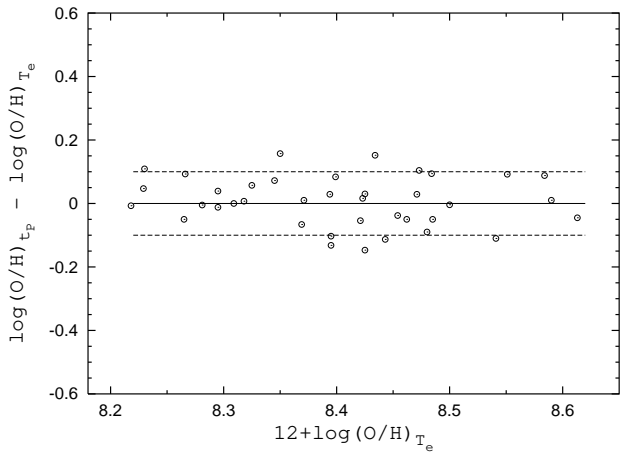}
\caption{\label{figure:10396f11}
The differences (O/H)$_{t_{P}}$ -- (O/H)$_{T_{e}}$ for our set of HII regions. 
}
\end{figure}

\begin{figure}[thb]
\vspace{6.5cm}
\includegraphics{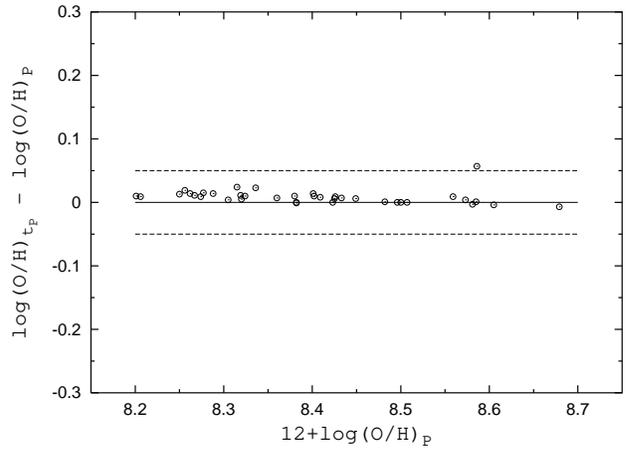}
\caption{\label{figure:10396f12}
The differences (O/H)$_{t_{P}}$  -- (O/H)$_{P}$ for our set of HII regions.
}
\end{figure}

Since the electron temperatures t$_{P}$ derived through the present calibration 
are close to the measured electron temperatures T$_{e}$, the t$_{P}$ can be
used for the oxygen abundance determination through the T$_{e}$ -- method. The 
oxygen abundances O/H$_{t_{P}}$ in HII regions from our sample determined in this 
way are given in Table \ref{table:data} (column 12). The differences between 
oxygen abundances derived with the t$_{P}$ and with the measured T$_{e}$ are 
presented in Fig.\ref{figure:10396f11}. The variant of the t$_{P}$=f(P,R$_{23}$) 
calibration which results in the best agreement between O/H$_{t_{P}}$ and 
O/H$_{T_{e}}$ for our set of HII regions and between O/H$_{t_{P}}$ and 
O/H$_{R_{G}}$ for low-excitation HII regions in the disk of our Galaxy was 
chosen above.

The differences between O/H$_{t_{P}}$ and O/H$_{P}$  for our set of HII regions 
are presented in Fig.\ref{figure:10396f12}. It can be seen in 
Fig.\ref{figure:10396f12} that O/H=f(P,R$_{23}$) and t$_{P}$=f(P,R$_{23}$) 
relations lead to consistent abundances.

Thus, the positions in the R$_{23}$ -- P diagram can be calibrated in terms of 
electron temperatures. The obtained relation t$_{P}$=f(P,R$_{23}$) between 
electron temperature and parameters P and R$_{23}$ provides an alternative 
(to the relation O/H=f(P,R$_{23}$) obtained in the previous section) method of 
the oxygen abundance determination in high-metallicity HII regions with 
the undetectable [OIII]4363 line. 

\section{Discussion}

The starting expression (Eq.(\ref{equation:kR3})) for the search for the relation 
$T_{e}$ = f(P,R$_{23}$) has been chosen from analysis of positions of HII
regions in the R$_{3}$ -- P (Fig.\ref{figure:10396f04}) and R$_{23}$ -- P 
(Fig.\ref{figure:10396f05}) diagrams. These diagrams are based on directly measured 
values. The starting Eq.(\ref{equation:kR3}) can be rewritten in the form
\begin{equation}
k_{0}R_{23} = k_{1} + k_{2} P + k_{3} P^{2},
\label{equation:tR23}
\end{equation}
where coefficients k$_{j}$ are dependent on the electron temperature.
With our assumption that the parameter P is an indicator of the hardness of 
the ionizing radiation, Eq.(\ref{equation:tR23}) is the relation between 
the hardness of the ionizing radiation and electron temperature in the nebula. 
On the other hand, it is well known that the relation between the hardness of 
the ionizing radiation (or effective temperature of the exciting star) and 
electron temperature in the nebula can be derived from the law of
energy conservation for free electrons. According to Sobolev (1967, Eq.23.37)
this relation is given by the expression
\begin{equation}
A\,T_{eff} = B\,T_{e} + C\,R_{23} + D \left\langle \frac{n_{1}}{n^{+}} \right\rangle
\label{equation:sobolev}
\end{equation}
where
\begin{equation}
\left\langle \frac{n_{1}}{n^{+}} \right\rangle  = 
\frac{\int n_{1}\,n_{e}\,dV}{\int n^{+}\,n_{e}\,dV} ,
\label{equation:sobolev2}
\end{equation}
where coefficient A depends on T$_{eff}$ only and B, C and D depend on T$_{e}$ 
only. Comparison of Eq.(\ref{equation:sobolev}) with Eq.(\ref{equation:tR23}) 
shows that with our assumption that the parameter P is an indicator of the 
hardness of the ionizing radiation (or effective temperature of the exciting 
star) our starting Eq.(\ref{equation:tR23}) is in some sense similar to 
Eq.(\ref{equation:sobolev}). Then, the relation $T_{e}$ = f(P,R$_{23}$) 
derived here can be considered as some kind analog of the equation of the energy 
balance of a gaseous nebula.

The validity of the obtained relation O/H=f(P,$R_{23}$) depends on the 
reliability of the oxygen abundances in HII regions for which this relation
has been derived. The two-zone model for T$_{e}$ is at the basis of the 
oxygen abundance determination in the present study. Then the O/H=f(P,$R_{23}$)
relation cannot be more credible than the two-zone model for T$_{e}$. If it is
ever established that the two-zone model for T$_{e}$ is a crude approximation 
of reality and leads to large uncertainty in the oxygen abundances then the 
oxygen abundances in HII regions should be redetermined within the framework 
of a more realistic model and O/H=f(P,$R_{23}$) relation should be revised.

Most metal-rich HII regions are objects with small values of R$_{23}$.
Although both the O/H$_{P}$ and the O/H$_{R_{23}}$ are maximum for those
HII regions, the O/H$_{P}$ for them are significantly lower (up to 0.5 dex) 
as compared to the O/H$_{R_{23}}$. It must be emphasized, however, 
that the O/H=f(P,$R_{23}$) relation is established on the basis of HII regions 
with R$_{23}$ larger than 2 (panel a1 in Fig.\ref{figure:10396f08}) and the 
validity of this relation in the case of HII regions with R$_{23}$ less than 2 
may be disputed. Then an additional information should be invoked in order 
to firmly establish how rich the most oxygen-rich HII regions are.

\section{Conclusions}

The problem of line intensity -- oxygen abundance calibration has
been investigated starting from the idea of McGaugh 
(1991) that the strong oxygen lines ($[OII] \lambda \lambda 3727, 3729$ and 
$[OIII] \lambda \lambda 4959, 5007$) contain the necessary information to 
determine accurate abundances in HII regions. The high-metallicity HII 
regions (12+logO/H $\geq$ 8.2, the upper branch of the O/H -- R$_{23}$ diagram) 
are considered in the present study. The low-metallicity HII regions 
(12+logO/H $\leq$ 7.95, the lower branch of the O/H -- R$_{23}$ diagram) have 
been considered in the previous study (Pilyugin 2000).

An important proposition that in high-metallicity HII regions the value of 
the abundance index R$_{23}$ is mainly governed by the oxygen abundance 
and by the hardness of the ionizing radiation (or by effective temperature 
of the exciting star(s)) and depends very weakly (if at all) on the ionization 
parameter, is immediately evident from the following observational facts:
{\it 1)} the value of R$_{23}$ is relatively constant within a given HII region,
{\it 2)} there is no one-to-one correspondence between R$_{23}$ and oxygen 
abundance. Another important fact supported the use of the excitation parameter 
P as an indicator of the hardness of the ionizing radiation is also evident 
from the observational data. These propositions are at the basis of the strong 
oxygen line intensity -- oxygen abundance calibration.

A relation of the type O/H=f(P,$R_{23}$) was derived empirically using the 
available oxygen abundances determined via measurement of temperature-sensivite 
line ratios ($T_{e}$ -- method). By comparing oxygen abundances in HII regions 
derived with the $T_{e}$ -- method and those derived with the suggested relations 
(P -- method) it was found that the precision of oxygen abundance 
determination with the P -- method is around 0.1 dex and is comparable to that 
obtained with the $T_{e}$ -- method.

A relation of the type T$_{e}$=f(P,$R_{23}$) was derived empirically 
using the available electron temperatures determined via measurement of the
temperature-sensivite line ratio [OIII]4959,5007/[OIII]4363. The maximum value 
of differences between electron temperatures derived through the 
T$_{e}$=f(P,$R_{23}$) relation and determined via measurement of the  
temperature-sensitive line ratio is around 1000K for HII regions considered here; 
the mean difference is $\sim$ 500K, which is of the same order of magnitude as the 
uncertainties of electron temperature determinations in high-metallicity HII 
regions via measured temperature-sensivite line ratios themselves.

The relation T$_{e}$=f(P,R$_{23}$) between electron temperature and parameters 
P and R$_{23}$ provides an alternative (to the relation O/H=f(P,R$_{23}$)) 
method of oxygen abundance determination in high-metallicity HII regions with 
the undetectable [OIII]4363 line. By comparing oxygen abundances in HII regions 
derived through the $T_{e}$ -- method with measured electron temperatures and 
those derived through the $T_{e}$ -- method with electron temperature 
determined with help of Te=f(P,$R_{23}$) relation, it was found that the 
precision of oxygen abundance determination with the latter method is around 0.1 
dex and is comparable to that obtained with the $T_{e}$ -- method.

\begin{acknowledgements}
I thank the referee, Prof. B.E.J.Pagel, for helpful comments and suggestions
as well as improving the English text. This study was partly supported by the 
NATO grant PST.CLG.976036 and the Joint Research Project between Eastern Europe 
and Switzerland (SCOPE) No. 7UKPJ62178.
\end{acknowledgements}

\end{document}